\documentclass[12pt]{iopart}
\usepackage{iopams}
\usepackage{cite}
\usepackage{mathrsfs}
\usepackage{graphicx}

\begin{document}

\title[Dispersion relation in strongly a magnetized plasma]{Dispersion relation for electromagnetic wave propagation in a strongly magnetized plasma}
\author{G. Brodin M. Marklund, L. Stenflo and P.K. Shukla}
\address{Department of Physics, Ume{\aa} University, SE--901 87 Ume{\aa},
  Sweden}

\begin{abstract}
A dispersion relation for electromagnetic wave propagation in a strongly
magnetized cold plasma is deduced, taking photon--photon scattering into
account. It is shown that the combined plasma and quantum electrodynamic effect is important
for understanding the mode-structures in magnetar and pulsar atmospheres.
The implications of our results are discussed.
\end{abstract}


\section{Introduction}

The quantum electrodynamical (QED) phenomenon of elastic photon--photon
scattering, due to the interaction of photons with virtual
electron--positron pairs, has recently received increased attention 
\cite{Ding1992,Moulin1999,BMS2001,Eriksson2004,Shen-etal,Shen-Yu,Bulanov-etal,JPP,%
Marklund-Brodin-Stenflo,MSSBS2005,RMP}. 
Several papers are motivated by the desire to detect
photon--photon scattering in laboratories \cite
{Ding1992,Moulin1999,BMS2001,Eriksson2004}, whereas others \cite
{Shen-etal,Shen-Yu,Bulanov-etal} concern phenomena that might be relevant
when the laser power is further increased to produce electric fields
strengths close to the Schwinger field $\sim 10^{18}\, \mathrm{V\,m^{-1}}$ \cite{Bulanov-etal}%
. Up to now, however, observable effects of photon--photon scattering are
likely to occur only for astrophysical systems \cite
{Baring-Harding,Shaviv-etal,Marklund-Brodin-Stenflo,magnetar,Bialynicki-Birula1970,%
Adler,MSSBS2005}, where the
large magnetic field strengths in pulsar and magnetar environments \cite{magnetar,Duncan-Thompson,Palmer-etal}
open up
for QED processes to play an important role, leading to phenomena such as
frequency down-shifting \cite{Adler,Bialynicki-Birula1970} and lensing 
\cite{Shaviv-etal}. \ The
frequency down-shifting \ is a result of so called photon splitting \cite
{Adler,Bialynicki-Birula1970}, which is one of the consequences of elastic
photon-photon scattering, and the process may even be responsible for the
radio silence of magnetars \cite{Baring-Harding}. Another QED-process of interest
in pulsar and magnetar environments is pair-production \cite{Beskin-book}
due to the strong field interactions, which lead to the presence of an
electron-positron pair plasma in the pulsar and magnetar atmospheres.
However, with a few exceptions (e.g.\ \cite{MSSBS2005,JPP}), we note that
most papers considering photon--photon scattering have omitted plasma effects
when considering electromagnetic wave propagation under these conditions.

In the present paper we will consider electromagnetic wave propagation at an
arbitrary angle to a strong external magnetic field $\mathbf{B}_{0}$, and
include the QED-effects associated with that field, as well as the influence
of an electron-positron pair plasma. The former effect is described within
the framework of the Heisenberg--Euler Lagrangian, which constitutes an
effective theory of photon--photon scattering \cite
{Heisenberg-Euler,Schwinger}, and the latter contribution follows from
elementary plasma theory. A comparatively general dispersion relation will
be derived. It reduces to previous results in a number of limiting cases 
\cite{Chen-book,Bialynicki-Birula1970,MSSBS2005}. In order to determine the
contribution from the pair-plasma on the propagation properties in pulsar
and magnetar atmospheres, we adopt the Goldreich-Julian expression for the
plasma density \cite{Beskin-book}, and evaluate the dispersion relation for
field strengths in the pulsar and magnetar range, $B_{0}\sim 10^{8}-10^{10}%
\,\mathrm{T}$. In the radio-wave regime it then turns out that for one of the
EM-wave polarizations, the plasma effects are typically negligible as
compared to the QED-effects, whereas for the other polarization, the
opposite is true in most cases. Noting that important processes in pulsar
and magnetar environments, e.g.\ photon splitting, typically involve both
EM-wave polarizations, we will conclude that QED and plasma effects should
be simultaneously included when studying radio wave propagation in such
environments.

\section{Derivations}

If vacuum fluctuations are taken into account, such as under highly
energetic conditions (e.g.\ pulsar plasmas and the next generation of
laser-plasma systems), Maxwell's equation will be altered by the quantum
vacuum self-interaction through the polarization 
\begin{equation}
\mathbf{P}=2\kappa \epsilon _{0}^{2}\left[ 2(E^{2}-c^{2}B^{2})\mathbf{E}%
+7c^{2}(\mathbf{E}\cdot \mathbf{B})\mathbf{B}\right]
\end{equation}
and magnetization 
\begin{equation}
\mathbf{M}=2\kappa \epsilon _{0}^{2}c^{2}\left[ -2(E^{2}-c^{2}B^{2})\mathbf{B%
}+7(\mathbf{E}\cdot \mathbf{B})\mathbf{E}\right]
\end{equation}
respectively, see e.g.\ \cite{BMS2001}. Here $\kappa =(\alpha /90\pi
)(1/\epsilon _{0}E_{\mathrm{crit}}^{2})$ gives the strength of the quantum
vacuum nonlinearity, $\alpha \approx 1/137$ is the fine-structure constant, $%
E_{\mathrm{crit}}=m^{2}c^{3}/e\hbar \sim 10^{18}\,\mathrm{V\,m^{-1}}$ is the
Schwinger critical field, $m$ is the electron rest mass, $c$ is the speed of
light in vacuum, $e$ is the magnitude of the electron charge, and $\hbar $
is Planck's constant divided by $2\pi $. These corrections to Maxwell's
vacuum equations are valid as long as $|\mathbf{E}|\ll E_{\mathrm{crit}}$ and $%
\omega \ll \omega _{e}=mc^{2}/\hbar $, where $\omega _{e}\approx 8\times
10^{20}\,\mathrm{rad\,s^{-1}}$ is the Compton frequency.

Next we Fourier decompose the electromagnetic perturbations, which have
frequencies $\omega $ and wavevectors $\mathbf{k}$. Maxwell's equations
together with the plasma equations of motion then yield 
\begin{equation}
\Delta ^{ab}\delta E_{b}=0.
\end{equation}
using index notation. Here the matrix $\Delta ^{ab}=n^{a}n^{b}-n^{2}\delta
^{ab}+\epsilon ^{ab}$, where $n^{a}=k^{a}c/\omega $, $n=kc/\omega $ is the
plasma refractive index, where $k=|\mathbf{k}|$, $\epsilon ^{ab}=\epsilon _{%
\mathrm{classical}}^{ab}+\epsilon _{\mathrm{QED}}^{ab}$ is the dielectric
tensor, 
\begin{equation}
\epsilon _{\mathrm{classical}}^{ab}=\delta ^{ab}+i\omega \sum_{s}\left( \frac{%
\omega _{\mathrm{p}s}}{\omega }\right) ^{2}\sigma _{s}^{ab},
\label{eq:dielectric-classical}
\end{equation}
\begin{equation}
\epsilon _{\mathrm{QED}}^{ab}=-4\xi \left[ \delta ^{ab}+n^{a}n^{b}-n^{2}\delta
^{ab}-\frac{7}{2}b^{a}b^{b}-2(\eta ^{aij}n_{i}b_{j})(\eta ^{bkl}n_{k}b_{l})%
\right] ,
\end{equation}
$s$ denotes the plasma particle species, $\omega
_{\mathrm{p}s}=(q_{s}^{2}n_{s}/\epsilon _{0}m_{s})^{1/2}$ is the plasma frequency for
species $s$, $\ \xi =\kappa \epsilon _{0}c^{2}B_{0}^{2}=(\alpha /90\pi
)(cB_{0}/E_{\mathrm{crit}})^{2}$ is the dimensionless QED parameter, $%
b^{a}=B_{0}^{a}/B_{0}$, and 
\begin{equation}
(\sigma _{s}^{ab})^{-1}=-i\omega \delta ^{ab}+\omega _{\mathrm{c}s}\eta ^{abj}b_{j},
\label{eq:sigmainv}
\end{equation}
with the cyclotron frequency $\omega _{\mathrm{c}s}=q_{s}B_{0}/m_{s}$ for species $s,$
$\delta ^{ab}$ is the Kronecker dela and $\eta _{abc}$ is the totally
anti-symmetric unit tensor. From the definition (\ref{eq:sigmainv}) we
obtain 
\begin{equation}
\sigma _{s}^{ab}=\frac{i\omega }{\omega ^{2}-\omega _{\mathrm{c}s}^{2}}(\delta
^{ab}-b^{a}b^{b})-\frac{\omega _{\mathrm{c}s}}{\omega ^{2}-\omega _{\mathrm{c}s}^{2}}\eta
^{abj}b_{j}+\frac{i}{\omega }b^{a}b^{b},  \label{sigma-QED}
\end{equation}
and the dielectric tensor (\ref{eq:dielectric-classical}) is thus 
\begin{equation}
\fl \epsilon _{\mathrm{classical}}^{ab}=\delta ^{ab}-\sum_{s}\left[ \frac{\omega
_{\mathrm{p}s}^{2}}{\omega ^{2}-\omega _{\mathrm{c}s}^{2}}(\delta ^{ab}-b^{a}b^{b})+\frac{%
i\omega _{\mathrm{p}s}^{2}\omega _{\mathrm{c}s}}{\omega (\omega ^{2}-\omega _{\mathrm{c}s}^{2})}\eta
^{abj}b_{j}+\left( \frac{\omega _{\mathrm{p}s}}{\omega }\right) ^{2}b^{a}b^{b}\right]
,  \label{Eps-classic}
\end{equation}
We note that the full dielectric tensor depends on the wavevector through
the QED contribution $\epsilon _{\mathrm{QED}}^{ab}$. Freely propagating waves
are characterized by the vanishing of the dispersion relation $D(\omega ,%
\mathbf{k})=\mathrm{det}(\Delta ^{ab})$. Writing the QED-tensor $\epsilon _{%
\mathrm{QED}}^{ab}$in matrix form, letting the $\mathbf{k}$-vector lie in
the $xz$-plane, we then obtain 
\begin{equation}
\epsilon^{ab}_{\mathrm{QED}}=-4\xi \left( 
\begin{array}{ccc}
1-n_{\Vert }^{2} & 0 & n_{\bot }n_{\Vert } \\ 
0 & 1-n^{2}-2n_{\bot }^{2} & 0 \\ 
n_{\bot }n_{\Vert } & 0 & -\frac{5}{2}-n_{\bot }^{2}
\end{array}
\right)   \label{QED-matrix}
\end{equation}
where $n_{\Vert }=k_{\Vert }c/\omega $, $n_{\bot }=k_{\bot }c/\omega $ and
the $\mathbf{k}$-vector is written as $\mathbf{k}=k_{\bot }\widehat{\mathbf{x%
}}+k_{\Vert }\widehat{\mathbf{z}}$. From (\ref{Eps-classic}) the classical
contributions to $\Delta ^{ab}$ is 
\begin{equation}
\!\!\!\!\!\!\! \left( 
\begin{array}{ccc}
1- \displaystyle{\sum\limits_{s}\frac{\omega _{\mathrm{p}s}^{2}}{\omega ^{2}-\omega _{\mathrm{c}s}^{2}} }%
-n_{\Vert }^{2} & \displaystyle{i\sum\limits_{s}\frac{\omega _{\mathrm{p}s}^{2}\omega _{\mathrm{c}s}}{\omega
(\omega ^{2}-\omega _{\mathrm{c}s}^{2})}} & n_{\bot }n_{\Vert } \\ 
\displaystyle{-i\sum\limits_{s}\frac{\omega _{\mathrm{p}s}^{2}\omega _{\mathrm{c}s}}{\omega (\omega
^{2}-\omega _{\mathrm{c}s}^{2})}} & 1-\displaystyle{\sum\limits_{s}\frac{\omega _{\mathrm{p}s}^{2}}{\omega
^{2}-\omega _{\mathrm{c}s}^{2}}}-n^{2} & 0 \\ 
n_{\bot }n_{\Vert } & 0 & 1-\displaystyle{\sum\limits_{s}\frac{\omega _{\mathrm{p}s}^{2}}{\omega
^{2}}}-n_{\bot }^{2} 
\end{array}
\right)   \label{Classical-matrix}
\end{equation}
The determinant of the sum of the matrixes (\ref{QED-matrix}) and (\ref
{Classical-matrix}) is then evaluated to give the dispersion relation 
\begin{eqnarray}
  \fl 
  0 =\left( (1-n^{2})(1-4\xi )-\sum\limits_{s}\frac{\omega _{\mathrm{p}s}^{2}}{\omega
  ^{2}-\omega _{\mathrm{c}s}^{2}}+8\xi n_{\bot }^{2}\right) \times   
\nonumber \\ \fl
  \left[\! \left(\!\! (1-n_{\Vert }^{2})(1-4\xi )- \!\! \sum\limits_{s}\frac{\omega
  _{\mathrm{p}s}^{2}}{\omega ^{2}-\omega _{\mathrm{c}s}^{2}}\! \right) \!\! \left(\!\! 1+10\xi
  - \!\! \sum\limits_{s}\frac{\omega _{\mathrm{p}s}^{2}}{\omega ^{2}}-n_{\bot }^{2}\left(
  1-4\xi \right)\!\! \right)\!\! -n_{\bot }^{2}n_{\Vert }^{2}\left( 1-4\xi \right) \!
  \right] -  
\nonumber \\  \fl
\left( \sum\limits_{s}\frac{\omega _{\mathrm{p}s}^{2}\omega _{\mathrm{c}s}}{\omega (\omega
^{2}-\omega _{\mathrm{c}s}^{2})}\right) ^{2}\left( 1+10\xi -\sum\limits_{s}\frac{%
\omega _{\mathrm{p}s}^{2}}{\omega ^{2}}-n_{\bot }^{2}\left( 1-4\xi \right) \right) 
\label{Full-DR}
\end{eqnarray}
The dispersion relation (\ref{Full-DR}) is the main result of the present
paper. It describes wave propagation at any angle to the external magnetic
field in a multi-component plasma, and it includes the QED effects
associated with the external magnetic field. 
Thus, it applies to high frequency electromagnetic waves of any polarization, as well as electrostatic oscillations and low frequency waves, such as Alfv\'en waves. As a specific example of how the plasma dispersion relation is affected by the QED effects we consider the case of an electron--positron plasma with $\omega \sim \omega_{\mathrm{p}} \ll |\omega_{\mathrm{c}}|$. The dispersion relation relation for the ordinary mode propagating perpendicular to the background magnetic field, with strength $\sim 10^{10}\,\mathrm{T}$ is depicted in figure 1. 
A number of limiting cases of (%
\ref{Full-DR}) have previously appeared in the literature. First, neglecting
the QED-effects (i.e.\ letting $\xi \rightarrow 0$), we immediately
obtain the standard dispersion relation for a cold multi-component plasma
(see e.g.\ \cite{Chen-book}). Alternatively, letting $\omega
_{\mathrm{p}}\rightarrow 0$, we note that the dispersion relation depends on the
propagation angle relative to the magnetic field. Furthermore, we note that
the indicies of refraction depend on the polarization even without a plasma.
These QED-effects due to a strong external magnetic field are wellknown
(often referred to as ''birefringence of vacuum''). Our dispersion relation
in the limit $\omega _{\mathrm{p}}\rightarrow 0$ agrees with those of previous
works, see e.g.\ \cite{Adler,Bialynicki-Birula1970}. The combined contribution
from the QED-effects due to a strong magnetic field and a non-zero plasma
density have previously been considered \cite{MSSBS2005} in the limit of
parallel propagation and allowing for large amplitudes. Taking the limit of
a small wave amplitudes in the dispersion relation (11) of reference \cite
{MSSBS2005}, and letting $n_{\bot }\rightarrow 0$ in (\ref{Full-DR})
we obtain agreement with \cite{MSSBS2005}. 

\begin{figure}
\includegraphics[width=.9\textwidth]{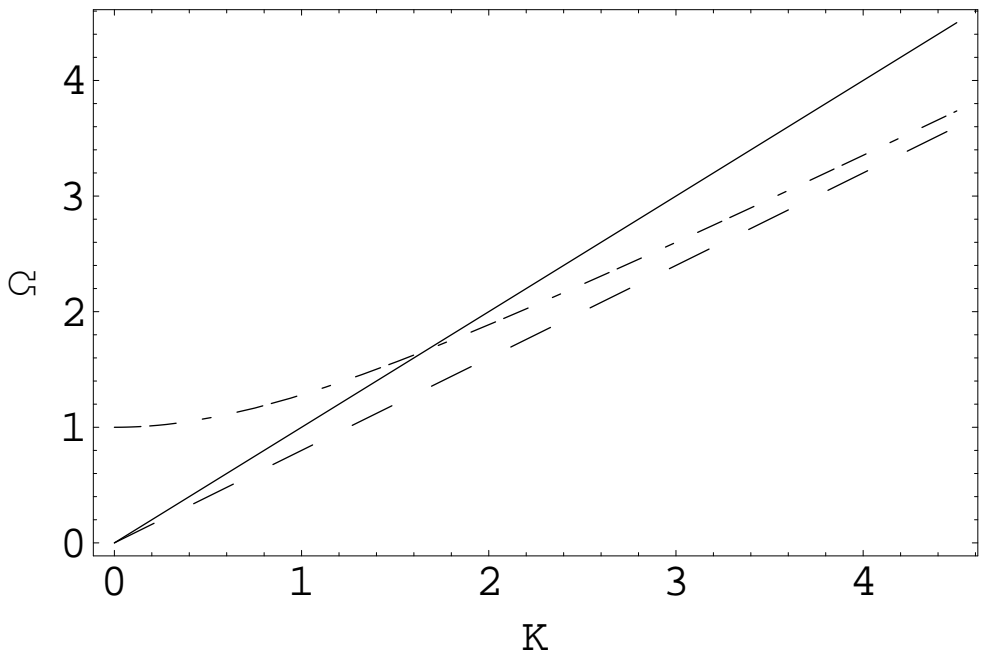}
\caption{The normalized frequency $\Omega = \omega/\omega_{\mathrm{p}}$ plotted as a function of the normalized wavenumber $K = ck/\omega_{\mathrm{p}}$ for three cases where $k_z = 0$. The full line represents the vacuum dispersion curve, the dashed line represents the strong magnetic field QED corrected vacuum dispersion, and the dashed-dotted line depicts the combined effect of plasma dispersion and QED effects. The QED corrections corresponds to a magnetic field strength of the order of $10^{10}\,\mathrm{T}$, i.e.\ field strengths expected around magnetars.}
\end{figure}

\section{Conclusion}

QED-effects associated with the external magnetic field are likely to be of
importance in environments with extreme magnetic fields, in particular in
the vicinity of astrophysical objects like pulsars and magnetars. For
example, the radio silence of magnetars is assumed to be connected with
QED-effects associated with the magnetar fields \cite{Baring-Harding}, which
could reach $10^{10} - 10^{11}\,\mathrm{T}$ $\ $\ close to the surface. However, in
the same environments, we also expect the presence of an electron-positron
plasma \cite{Beskin-book}. Thus we evaluate (\ref{Full-DR}) with $%
\sum_{s}=\sum_{e,p}$, where $e$ and $p$ denotes electrons and positrons,
respectively. Considering propagation at an arbitrary angle to the magnetic
field in an electron-positron plasma, letting $\omega _{\mathrm{p}e,\mathrm{p}p}\!\sim\! \omega\! \ll \! 
\left| \omega _{\mathrm{c}e,\mathrm{c}p}\right| $ , using $\xi\! \ll\! 1$ and noting that the
factor $[\sum_{e,p}\omega _{\mathrm{p}s}^{2}\omega _{\mathrm{c}s}/\omega (\omega ^{2}-\omega
_{\mathrm{c}s}^{2})]^{2}$ then becomes negligibly small (due to the approximate
cancellation of the electron and positron contributions), we find from (%
\ref{Full-DR}) that the dispersion relation separates in two modes that can
be approximated by 
\begin{equation}
1-n^{2}+8\xi n_{\bot }^{2}+\frac{\omega _{\mathrm{p}}^{2}}{\omega _{\mathrm{c}}^{2}(1-4\xi )}%
\approx 0  \label{Magnetosonic-DR}
\end{equation}
and 
\begin{equation}
(1-n^{2})(1-4\xi ) - \left( -14\xi +\frac{\omega _{\mathrm{p}}^{2}}{\omega ^{2}}%
\right) (1-n_{\Vert }^{2})  \approx 0 ,
\label{O-mode-DR}
\end{equation}
where $\omega _{\mathrm{p}}=(\omega _{\mathrm{p}e}^{2}+\omega _{\mathrm{p}p}^{2})^{1/2}$ is the total
plasma frequency, and $\omega _{\mathrm{c}}$ $=eB_{0}/m$ is the magnitude of the
electron (or positron) cyclotron frequency. In the vicinity of pulsars or
magnetars where $\omega _{\mathrm{c}}\sim 10^{19}\! -\! 10^{21}\,\mathrm{rad\,s^{-1},}$ the last term of (\ref{Magnetosonic-DR}) is negligible unless the plasma density is
extremely high. Omitting that term, the dispersion relation (\ref
{Magnetosonic-DR}) is then the same as that used in \cite
{Bialynicki-Birula1970} for the high phase velocity mode when considering
photon-splitting. Similarly the ordinary mode described by (\ref
{O-mode-DR}), reduces to the mode with the lower phase velocity of reference \cite
{Bialynicki-Birula1970} when the plasma is removed. However, for the latter
dispersion relation we note that a relatively modest plasma density is
enough to significantly affect the propagation properties in the radio wave
regime. To make a concrete estimate, we adopt the Goldreich--Julian density 
\begin{equation}
n_{\mathrm{GJ}}=7\times 10^{15}\left(\frac{0.1}{\tau }\right)\left(\frac{B_{\mathrm{pulsar}}}{10^{8}}\right)\,\mathrm{m}^{-3}
\label{Julian-Goldreich}
\end{equation}
where $\tau $ is the pulsar period time (in seconds) and $B_{\mathrm{pulsar}}$ the pulsar
magnetic field (in tesla). The pair plasma density is expected to satisfy $%
n_{e}=n_{p}=Mn_{\mathrm{GJ}}$, where $M$ is the multiplicity \cite
{Beskin-book,Luo-etal}. Moderate estimates then give $M=10$ \cite{Luo-etal}.
Choosing this value and letting $\tau =1\,\mathrm{s}$, we note that for
magnetar field strengths, $B_{\mathrm{pulsar}}=10^{10}\,\mathrm{T}$, the term due to the
plasma $\propto \omega _{\mathrm{p}}^{2}/\omega ^{2}$ in (\ref{O-mode-DR})
dominates over the term due to QED $\propto 14\xi $ for frequencies up to $%
\omega \sim 10^{14} - 10^{15}\,\mathrm{rad\,s^{-1},}$ i.e.\ in the infrared regime and
below. Furthermore, we note that photon splitting \cite
{Adler,Bialynicki-Birula1970} as described by standard QED (i.e.\ with zero plasma
density) requires that the phase velocity of the dispersion relation in (%
\ref{Magnetosonic-DR}) is higher than that of (\ref{O-mode-DR}). While
this is always true in the absence of a plasma, we note that for wave
frequencies in the infra-red regime and below, the Goldreich--Julian density
given by (\ref{Julian-Goldreich}) is enough to increase the phase velocity
of the mode in (\ref{O-mode-DR}) above that of (\ref{Magnetosonic-DR}),
unless we choose the period time $\tau $ extremely low. \ Thus we conclude
that photon--photon splitting as described by vacuum theories is not likely
to apply to magnetar atmospheres, unless the pair-production \cite
{Beskin-book} responsible for the Goldreich-Julian expression is effectively
suppressed. Wave cascade processes as a mechanism to explain the radio
silence of magnetars \cite{Baring-Harding} could still be possible, but for
densities of the order of (\ref{Julian-Goldreich}), plasma nonlinearities
are likely to dominate over the pure QED effects.

\end{document}